\documentclass[%
 reprint,
 showkeys,
 amsmath,amssymb,
 aps,
]{revtex4-2}
\usepackage{xcolor, soul}
\sethlcolor{yellow}
\usepackage{graphicx}
\usepackage{dcolumn}
\usepackage{bm}
\usepackage{lipsum}

\usepackage{subfigure}

\begin{document}


\title{
A short-circuited coplanar waveguide for low-temperature single-port ferromagnetic resonance spectroscopy set-up to probe the magnetic properties of ferromagnetic thin films\\
}

\author{Sayani Pal, Soumik Aon, Subhadip Manna and Chiranjib Mitra}
\email{Corresponding author:chiranjib@iiserkol.ac.in}
\affiliation{%
 Indian Institute of Science Education and Research Kolkata, West Bengal, India\\
 }


\begin{abstract}
	
A coplanar waveguide shorted in one end is proposed, designed, and implemented successfully to measure the properties of magnetic thin films as a part of the vector network analyser ferromagnetic resonance (VNA-FMR) spectroscopy set-up. Its simple structure, potential applications and easy installation inside the cryostat chamber made it advantageous especially for low-temperature measurements. It provides a wide band of frequencies in the gigahertz range essential for FMR measurements. Our spectroscopy set-up with short-circuited coplanar waveguide has been used to extract Gilbert damping coefficient and effective magnetization values for standard ferromagnetic thin films like Py and Co. The thickness and temperature dependent studies of those magnetic parameters have also been done here for the afore mentioned magnetic samples.

\end{abstract}

\pacs{Valid PACS appear here}
\maketitle


\section*{Introduction}

In recent years, extensive research on microwave magnetization dynamics in magnetic thin films\cite{1t,2t,3t}, planar nanostructures\cite{1n,2n,3n} and multi-layers \cite{1m,2m,3m} have been performed due to their potential applications in various fields of science and technology. Spintronics is one such emerging discipline that encompasses the interplay between magnetization dynamics and spin transport. It also includes fields like spin-transfer torque \cite{S1,S2,S3,S4}, direct and inverse spin hall effect \cite{Sh1,Sh2,Sh3,Ish1,Ish2}, spin pumping \cite{Sp1,Sp2} etc., which are crucial in industrial applications for developing devices like magnetic recording head\cite{MR}, magnetic tunnel junction(MTJ) sensors \cite{MTJ1,MTJ2}, magnetic memory devices \cite{MM1,MM2} and spin-torque devices \cite{Sd1,Sd2}. Thus exploring more about the static and dynamic properties of magnetic materials in itself is an interesting subject. Ferromagnetic resonance spectroscopy(FMR) is a very basic and well-understood technique that is used to study the magnetization dynamics of ferromagnets\cite{F1, F2, VV1}. Nowadays, most advanced FMR spectroscopy methods use a vector network analyzer (VNA)\cite{Claus_Bilzer, VV1} as the microwave source and detector. We have used VNA in our set-up too.\\
To determine the magnetic parameters of the ferromagnetic materials using the VNA-FMR spectroscopy, one needs to carry out the measurements at a wide range of frequencies. Since the microwave magnetic field in the coplanar waveguide (CPW) is parallel to the plane, it serves the purpose of exploring the magnetic properties of the concerned system over a broad frequency range in the GHz region. The advantage of using CPW in the spectroscopy system lies in the fact that we no longer need to remount samples at different waveguides or cavities for every other frequency measurements, which consumes a lot of time and effort in an experiment\cite{TC1, TC2}. Researchers design and use different types of CPW for various other purposes like micron-sized CPW in microwave-assisted magnetic recording; two-port CPW in antenna; shorted CPW in ultra-wideband bandpass-filter and permeability measurements \cite{CU1, A short-circuited CPW, CU2}. However, in broadband FMR spectroscopy two-port CPW jigs have most commonly been used till date.  Using two-port CPW in FMR spectroscopy, one gets absorption spectra in terms of the transmission coefficient of scattering parameters, and from there magnetic parameters of the samples can be determined. The use of two-port CPW in VNA-FMR can be replaced by one-port CPW where the reflection coefficient of scattering parameters of the FMR spectra can be used to determine the magnetic parameters of the sample. One port reflection geometry is a lot more convenient in terms of easy design, calibration, installation, and sample loading. This is especially true when the whole CPW arrangement is kept inside the cryostat chamber for low-temperature measurements and the system becomes very sensitive to vibration and any kind of magnetic contacts, one port CPW seems very convenient to operate rather than the two-port one. Previously, many have designed and used short-circuited CPW jigs for other purposes but to the best our knowledge it has not been used for low-temperature VNA-FMR spectroscopy measurements before.\\
In this work, we report the development of short-circuited CPW based low-temperature broadband VNA-FMR spectroscopy set-up to study the magnetic parameters of standard ferromagnetic samples. For measurements, we chose the permalloy(Py) thin films as ferromagnetic (FM) material which has greatly been used in research fields like spintronics and industrial applications due to its interesting magnetic properties like high permeability, large anisotropy magnetoresistance, low coercivity, and low magnetic anisotropy. We have also considered another standard magnetic thin film, Co of thickness 30nm as a standard for ascertaining the measurement accuracy. In our system, we swept the magnetic field keeping the frequencies constant, and got the FMR spectra for several frequencies. From there we found the variation of resonance fields and field linewidths with the resonance frequencies. We have used the linear fit for resonance frequencies vs field line-widths data to calculate the Gilbert damping coefficient($\alpha$). We fitted the set of resonance frequencies vs resonance fields data to the Kittel formula \cite{Kittel} to obtain the effective magnetization($4\pi M_{eff}$). Subsequently, we investigated the thickness and temperature-dependent studies of $4\pi M_{eff}$ and $\alpha$ for FM thin films of different thickness in the temperature range of 7.5K to 300K. To characterise the measurement set-up using short-circuited CPW, we compared the previous measurements in the literature with our results and there was a good agreement between the two\cite{A short-circuited CPW,Experimantal_investigation}.

\section*{EXPERIMENTAL DETAILS}
A short-circuited CPW has been designed and fabricated as a part of our low-temperature VNA-FMR spectroscopy set-up. To make the CPW we have used Rogers AD1000, a laminated PCB substrate with copper cladding on both sides of the dielectric. The thickness of the dielectric and the copper layer are 1.5 mm and 17.5 microns respectively and the dielectric constant of the substrate is 10.7. The main concern about the design of the CPW is to match its characteristic impedance with the impedance of the microwave transmission line connected to it. We have used the line calculator to calculate the dimensions of CPW. For a CPW with a characteristic impedance of 50 ohms, the line calculator calculated the width of the signal line and the gap to be 900 microns and 500 microns respectively. The fabrication is done using optical lithography which is described in detail in the literature\cite{SR1}.
\begin{figure}[h!]
\centering  		
\includegraphics[width=9.0cm,height=5cm]{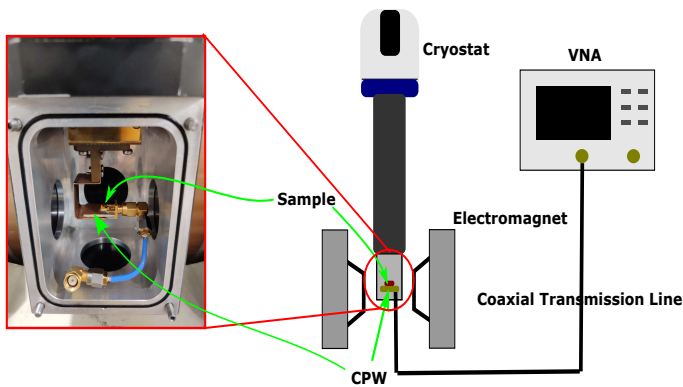}
\caption{The schematic diagram of measurement system and the arrangement inside the cryostat with the sample on top of the CPW}
\label{Fig 1}
\end{figure}
Other components of our measurement system are a)Vector Network Analyser(VNA), which is a microwave source as well as a detector, b)the electromagnet that generates the external magnetic field, i.e., Zeeman field and, c) optistat dry cryogen-free cooling system from Oxford instruments which is used for low-temperature measurements. One end of the CPW signal line is shorted to the ground, and the other end is connected to the VNA through a SMA connector and coaxial cable (fig 3b). On top of the CPW, thin-film samples have been placed face down after wrapping them with an insulating tape to electrically isolate them. For low-temperature measurements, the sample has been glued to the CPW using a low-temperature adhesive to ensure contact of sample and resonator at all times, in spite of the vibration caused by the cryostat unit. This whole arrangement is then placed inside the two pole pieces of the electromagnet as we can see from the diagram in fig 1. There are two standard methods of getting FMR spectra: sweeping the frequency keeping the field constant and sweeping the magnetic field while keeping the frequency constant. We have adopted the second method. We have worked in the frequency range from 2.5GHz  to 5.5GHz and in the magnetic field range from 0 Oe to roughly around 500 Oe. We have used 1mW of microwave power throughout the experiment. From the FMR spectra, we have determined effective magnetization and damping coefficient of FM thin films and studied their variation with temperature and sample thickness.

\section*{SAMPLE PREPARATION and CHARACTERIZATION}
\begin{figure}[h!]
\centering
\subfigure[]{\includegraphics[width=4.5cm,height=3cm]{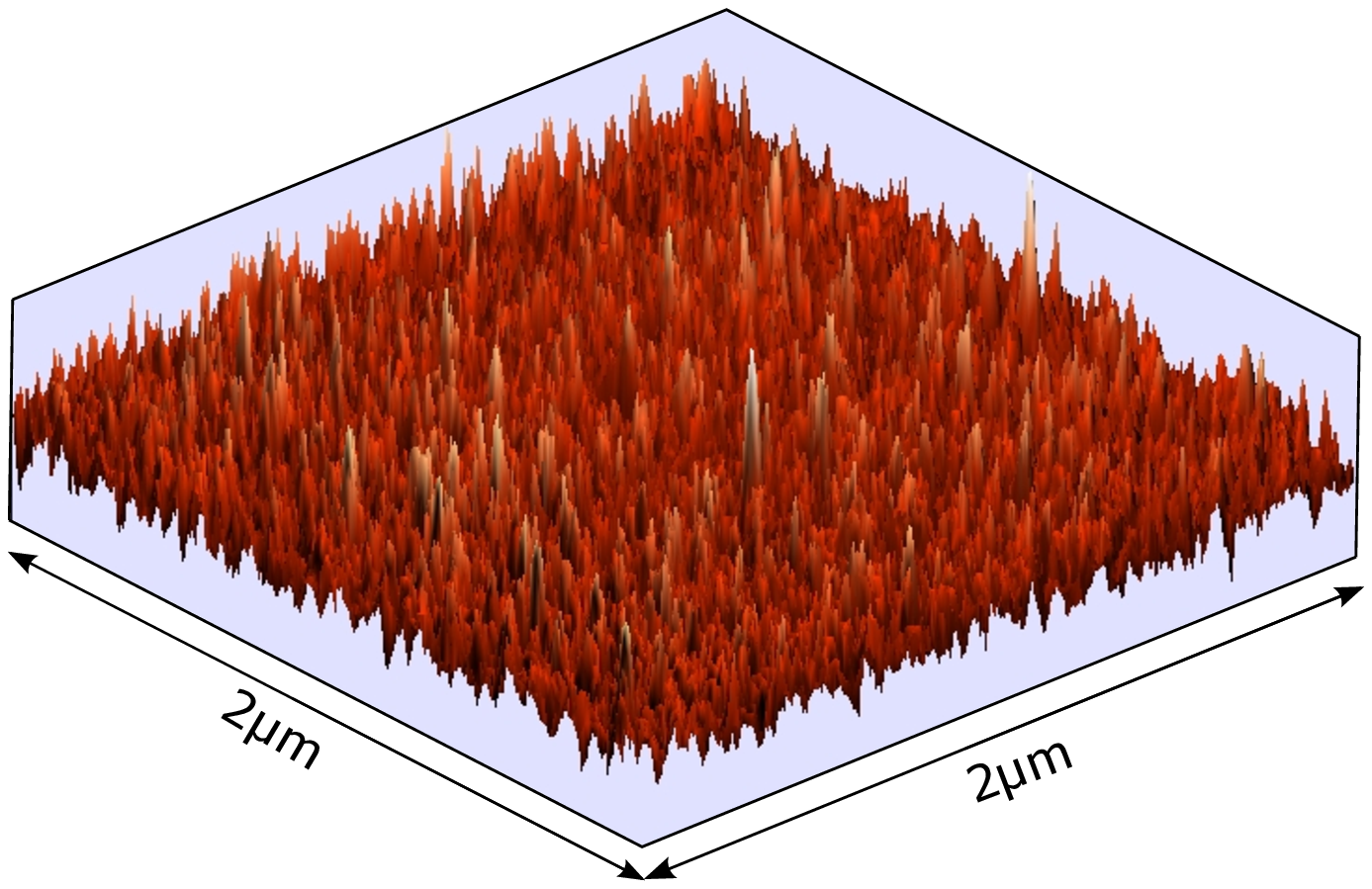}}\\
\subfigure[]{\includegraphics[width=6cm,height=5cm]{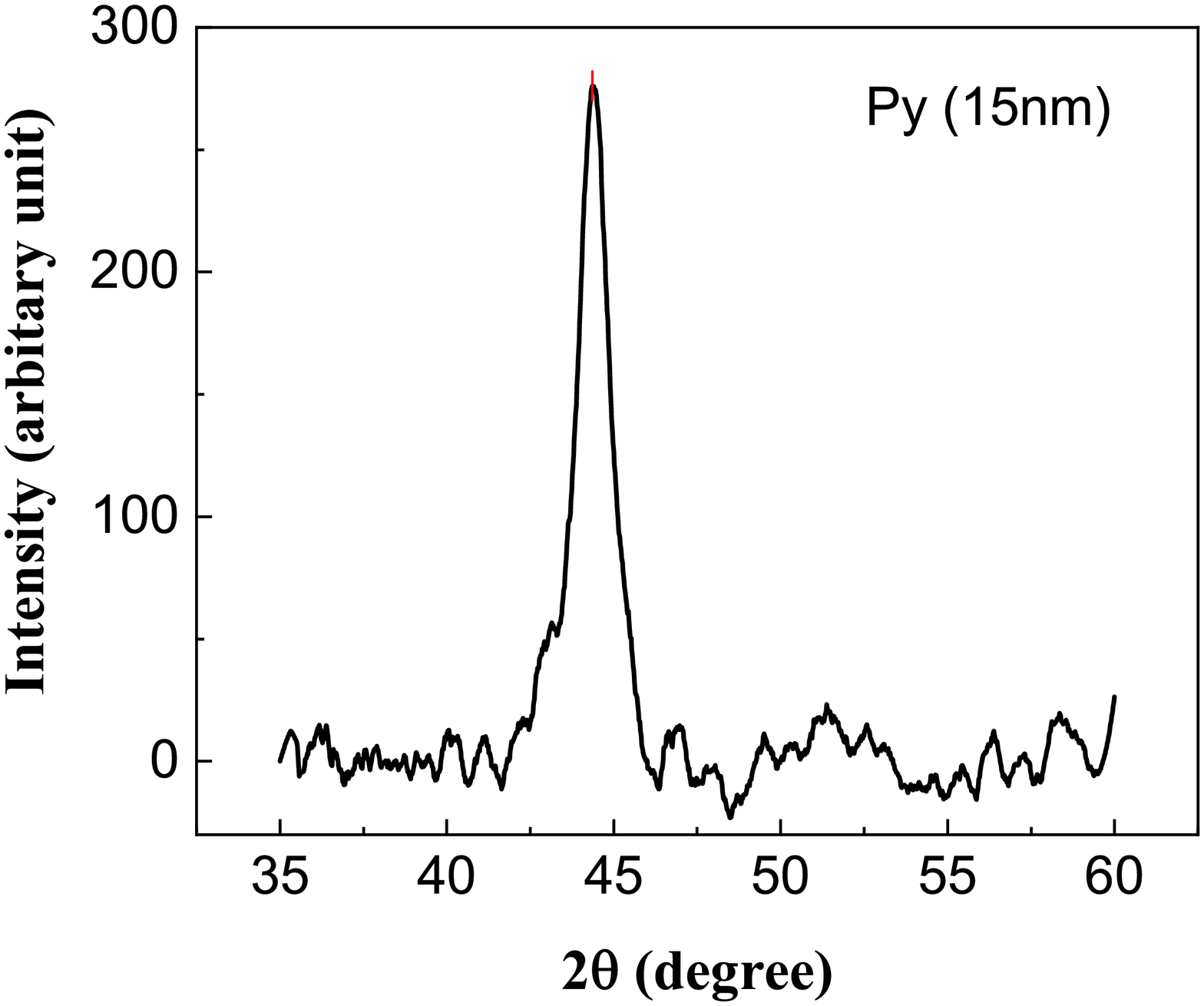}}
\subfigure[]{\includegraphics[width=6cm,height=5cm]{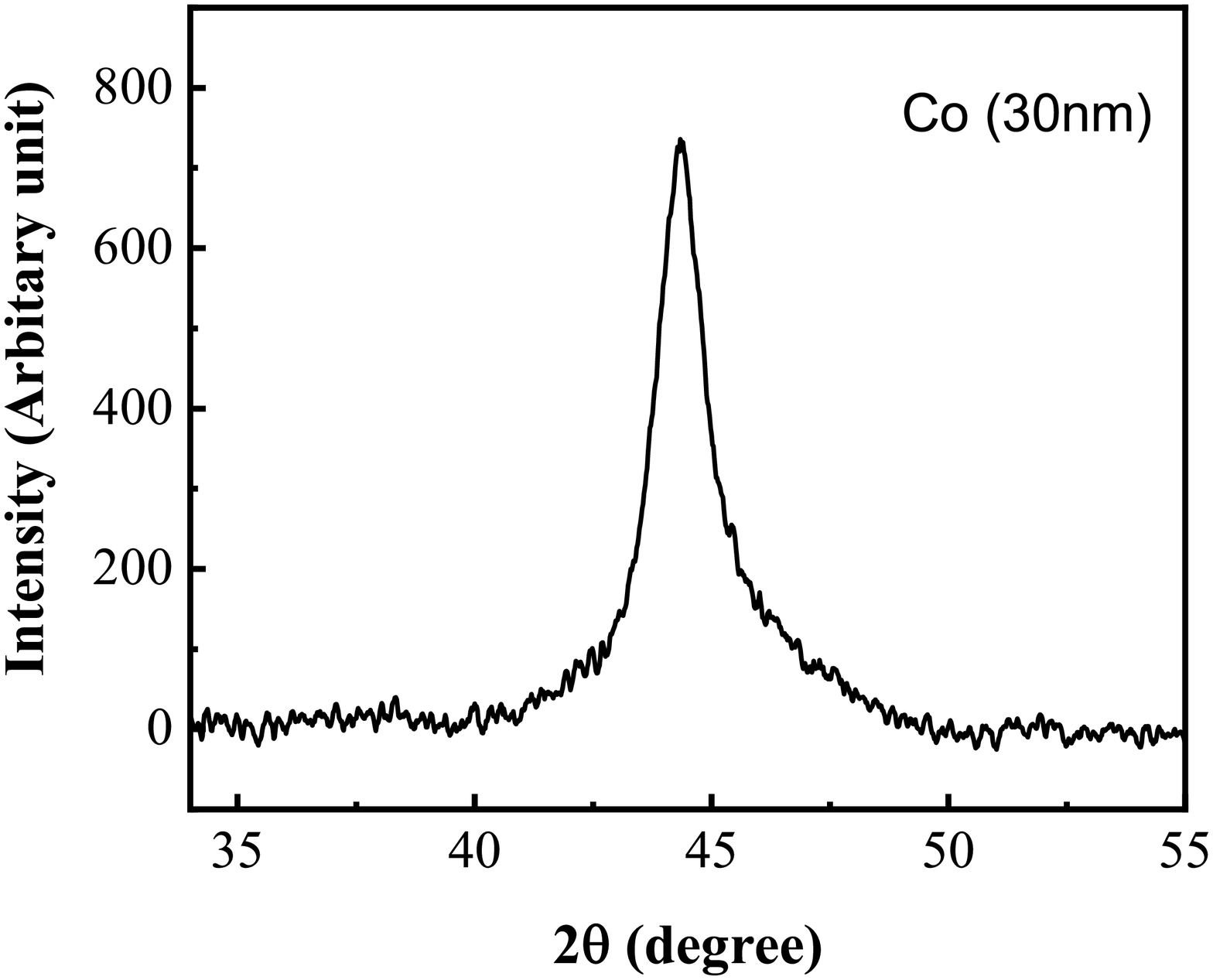}}
\caption{(a)Atomic force microscope (AFM) image of 30 nm thick Py thin film with a surface roughness of 1 nm . 
X-ray diffraction peak of (b)15nm thick Py film and (c)30nm Co prepared by thermal evaporation.}
\label{Fig 2}
\end{figure}

Py $(Ni_{80} Fe_{20})$ and Co thin films were fabricated by thermal evaporation technique on $Si/SiO_2$ substrates, from commercially available pellets $(99.995\% pure)$ at room temperature. The substrates were cleaned with acetone, IPA and DI water respectively in ultrasonicator and dried with a nitrogen gun. The chamber was pumped down to $1\times 10^{-7}$torr using a combination of a scroll pump and turbo pump. During the deposition, pressure reached upto $1\times10^{-6}$torr. Thin films were fabricated at a rate of $1.2 {\AA}/s$ where thickness can be controlled by Inficon SQM 160 crystal monitor. For our experiments a series of Py thin films of different thicknesses were fabricated by keeping the other parameters like base pressure, deposition pressure and growth rate constant. Film thickness and morphology was measured by using atomic force microscopy technique as shown in fig 2(a). We have used Py films with thicknesses 10nm, 15nm, 34nm, 50nm, and 90nm with a surface roughness of around 1nm and one Co film of thickness 30nm. X-ray diffraction experiment confirms the polycrystalline structure of the samples as shown in fig 2b and fig 2c for Py and Co respectively.

\section*{RESULTS and DISCUSSION}
We have calculated the dimensions of the short-circuited CPW using the line calculator of the CST Studio Suite software as mentioned in the experimental details section. Using those dimensions we have also done the full-wave electromagnetic simulation in CST software to get the electric and magnetic field distribution of the CPW. One can see from the simulation result displayed in figure 3a that the farther it is from the gap, the weaker the intensity of the magnetic field, and the magnitude of the field in the gap area is one order of greater than that on the signal line.
\begin{figure}[h!]
\centering  		
\includegraphics[width=7.0cm,height=10cm]{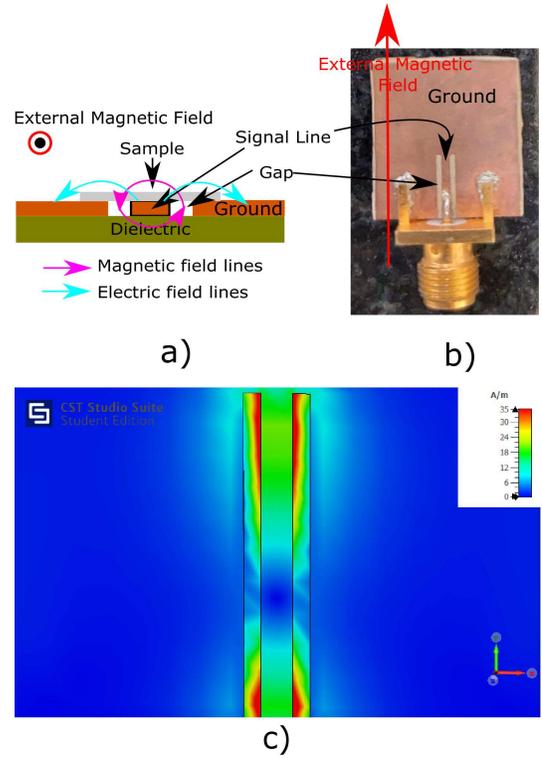}
\caption{(a) Schematic diagram of the cross-sectional view of CPW. (b) Top view of the short-circuited CPW after fabrication. (c)Intensity distribution of microwave magnetic field in the one end shorted CPW at 5GHz (top view)}
\label{Fig 3}
\end{figure}
\begin{figure}[h!]
\centering  		
\includegraphics[width=8.5cm,height=7cm]{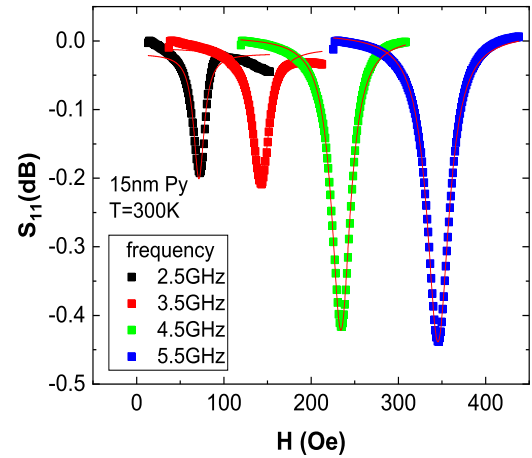}
\caption{Ferromagnetic Resonance spectra of absorption at frequencies 2.5 GHz, 3.5 GHz, 4.5 GHz, 5.5 GHz for 15nm Py thin films at room temperature after background subtraction}
\label{Fig 4}
\end{figure}
\begin{figure*}
\centering
\subfigure[]{\includegraphics[width=8cm,height=6cm]{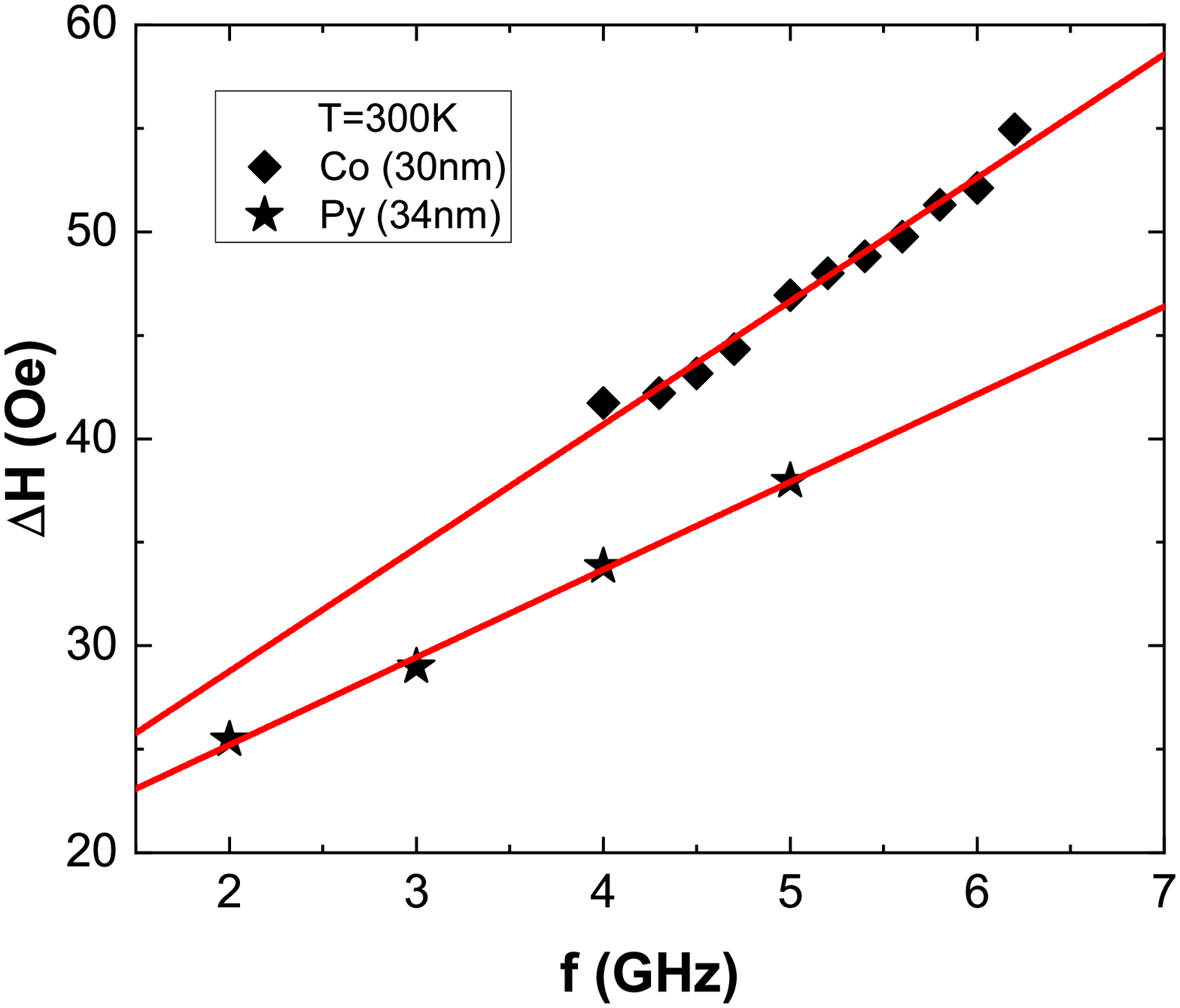}}
\subfigure[]{\includegraphics[width=8cm,height=6cm]{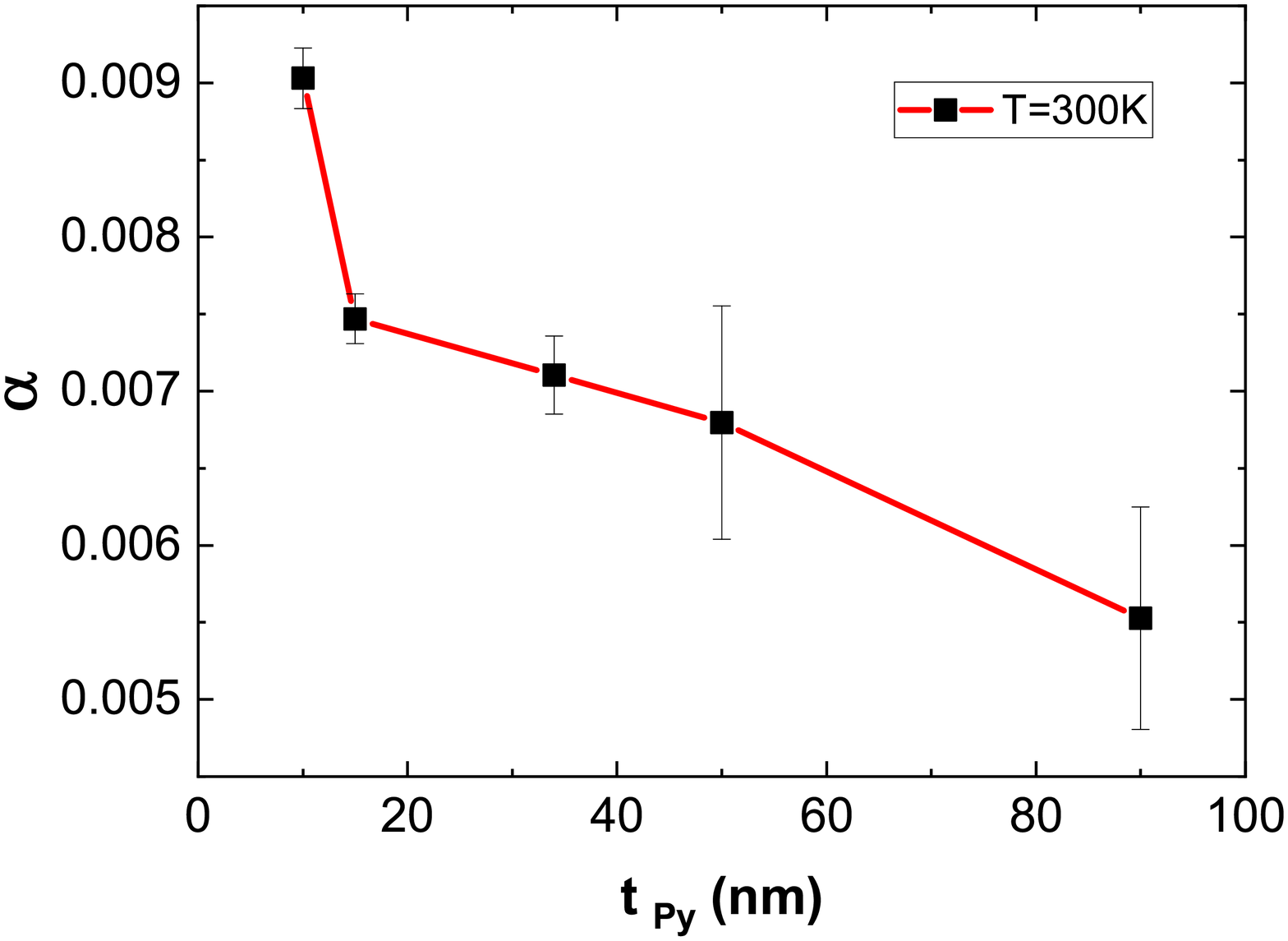}}
\hfill
\subfigure[]{\includegraphics[width=8cm,height=6cm]{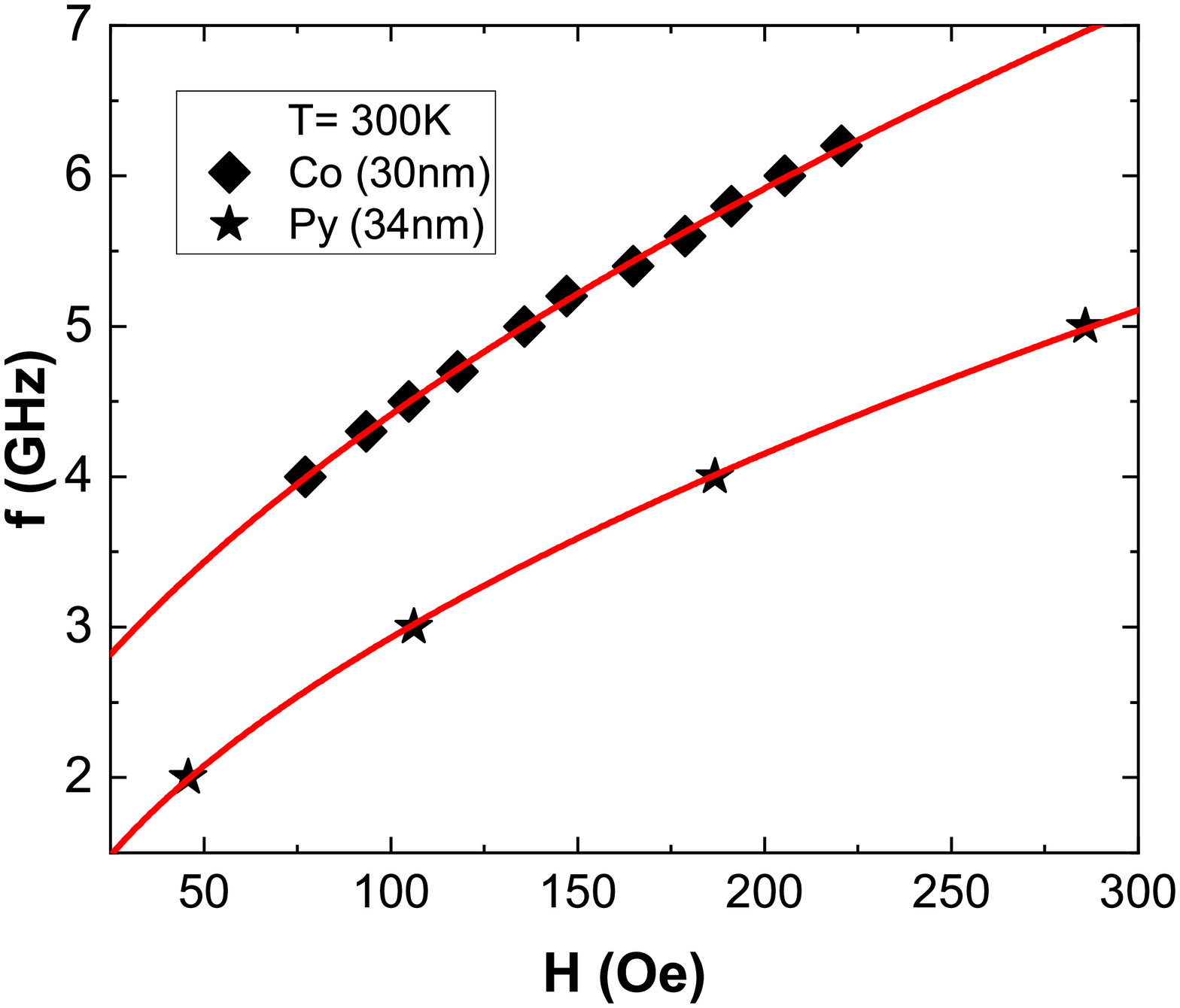}}
\subfigure[]{\includegraphics[width=8cm,height=6cm]{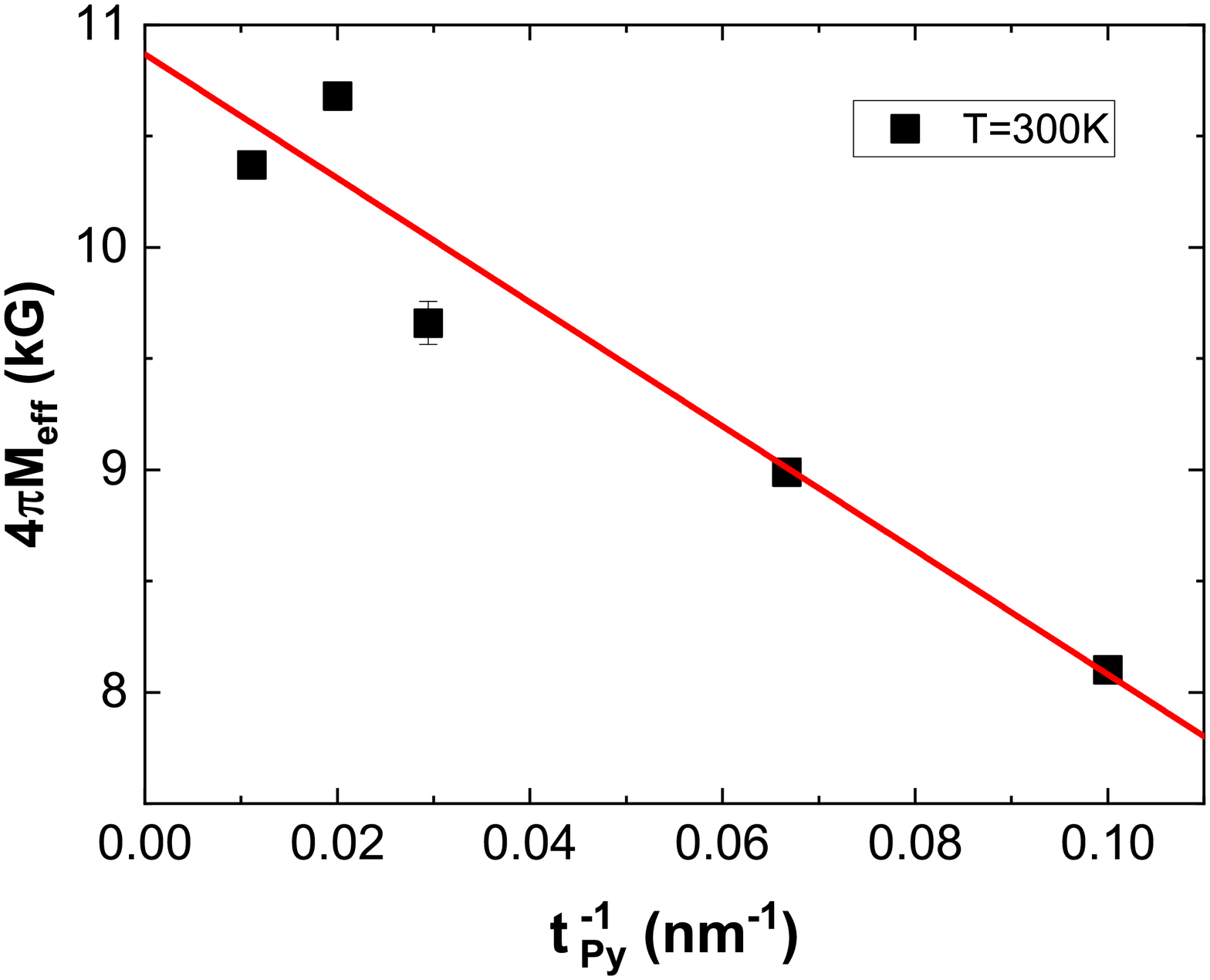}}
\caption{a)Field linewidth variation with resonance frequencies at 300K for 34nm Py and 30nm Co thin films. Equation 1 has been used for fitting the curve and to determine the Gilbert damping coefficient; b)thickness dependence of Gilbert damping coefficient at room temperature for Py thin films; c)resonance field variation with resonance frequencies at 300K for 34 nm Py and 30 nm Co thin films. Kittel formula (eqn-3) has been used for fitting the curve and to determine the effective magnetization; d)thickness dependence of effective magnetization for Py thin films at room temperature.}
\label{Fig 5}
\end{figure*}
When placing the thin film sample on top of the CPW, the dimension of the sample should be such that it can cover the gap area on both sides of the signal line of the CPW because the magnetic field is most intense in that area. This microwave magnetic field circulating the signal line of the CPW is perpendicular to the external magnetic field and both the magnetic fields are parallel to the film surface as can be seen from fig 3a and 3b. On account of the static magnetic field, the magnetic moment will undergo a precession around the static magnetic field at a frequency called the Larmor precession frequency. Absorption of electromagnetic energy happens when the frequency of the transverse magnetic field (microwave) is equal to the Larmor frequency.


\begin{figure*}
\centering
\subfigure[]{\includegraphics[width=8cm,height=6cm]{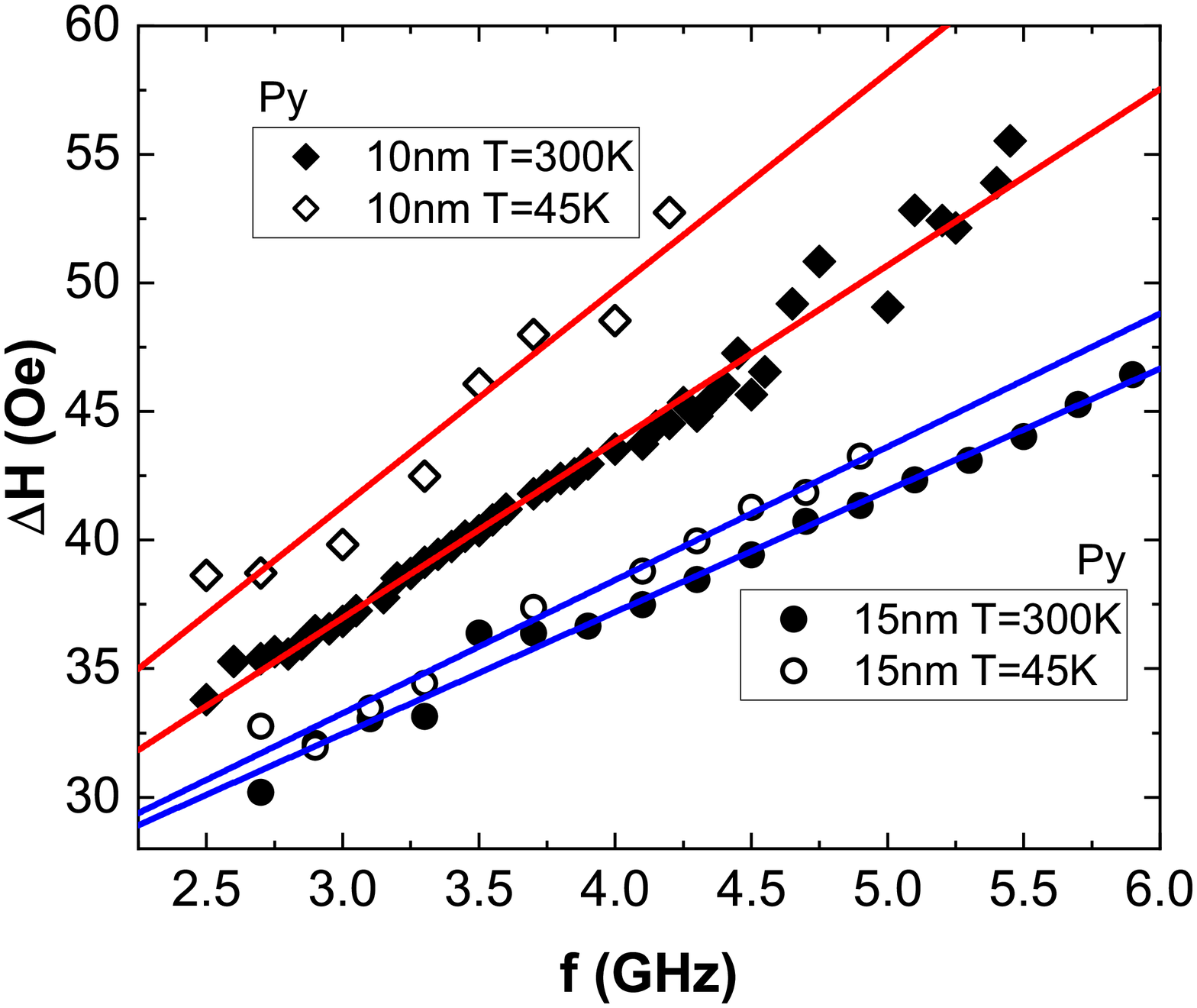}}
\subfigure[]{\includegraphics[width=8cm,height=6cm]{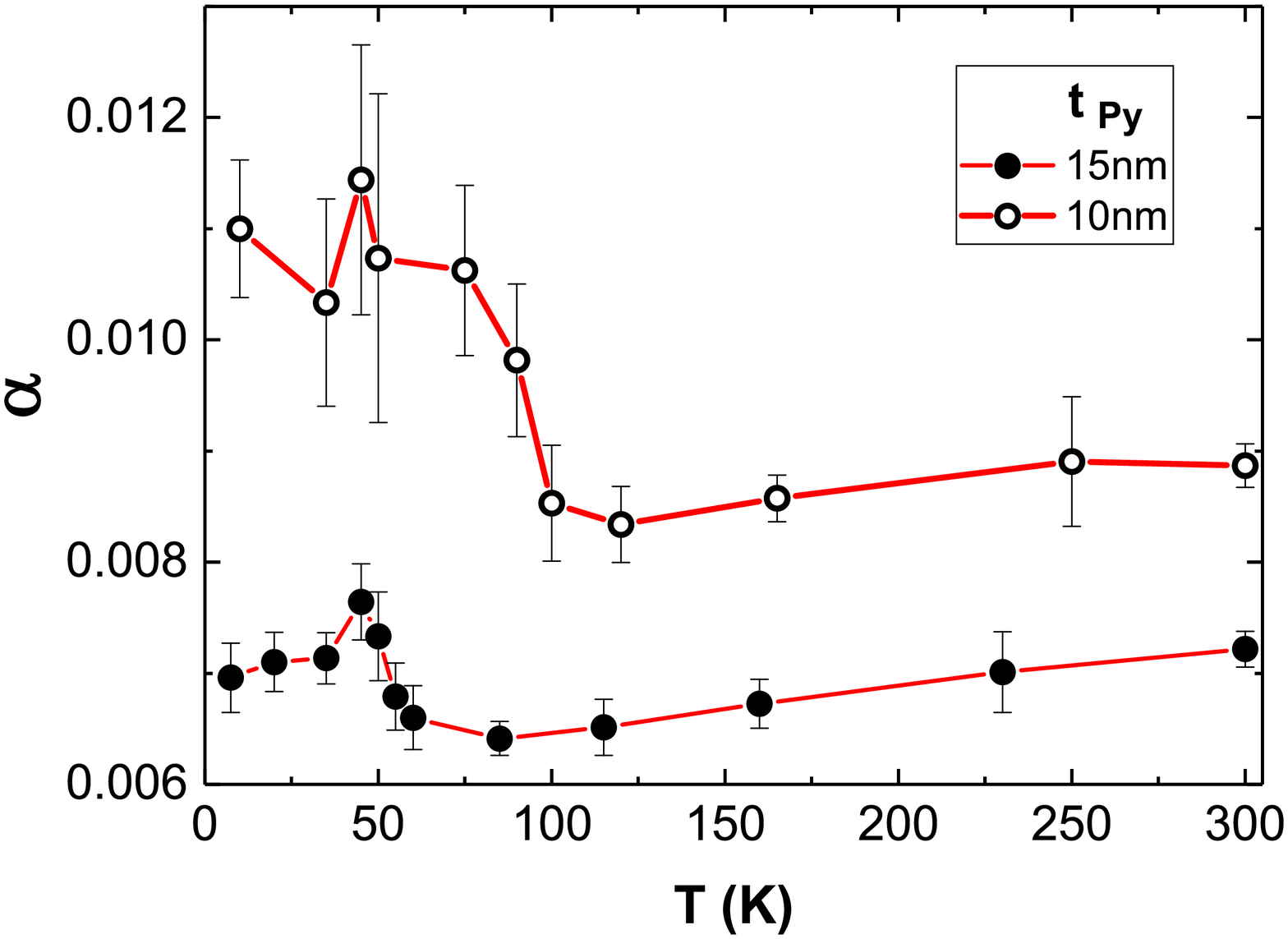}}
\hfill
\subfigure[]{\includegraphics[width=8cm,height=6cm]{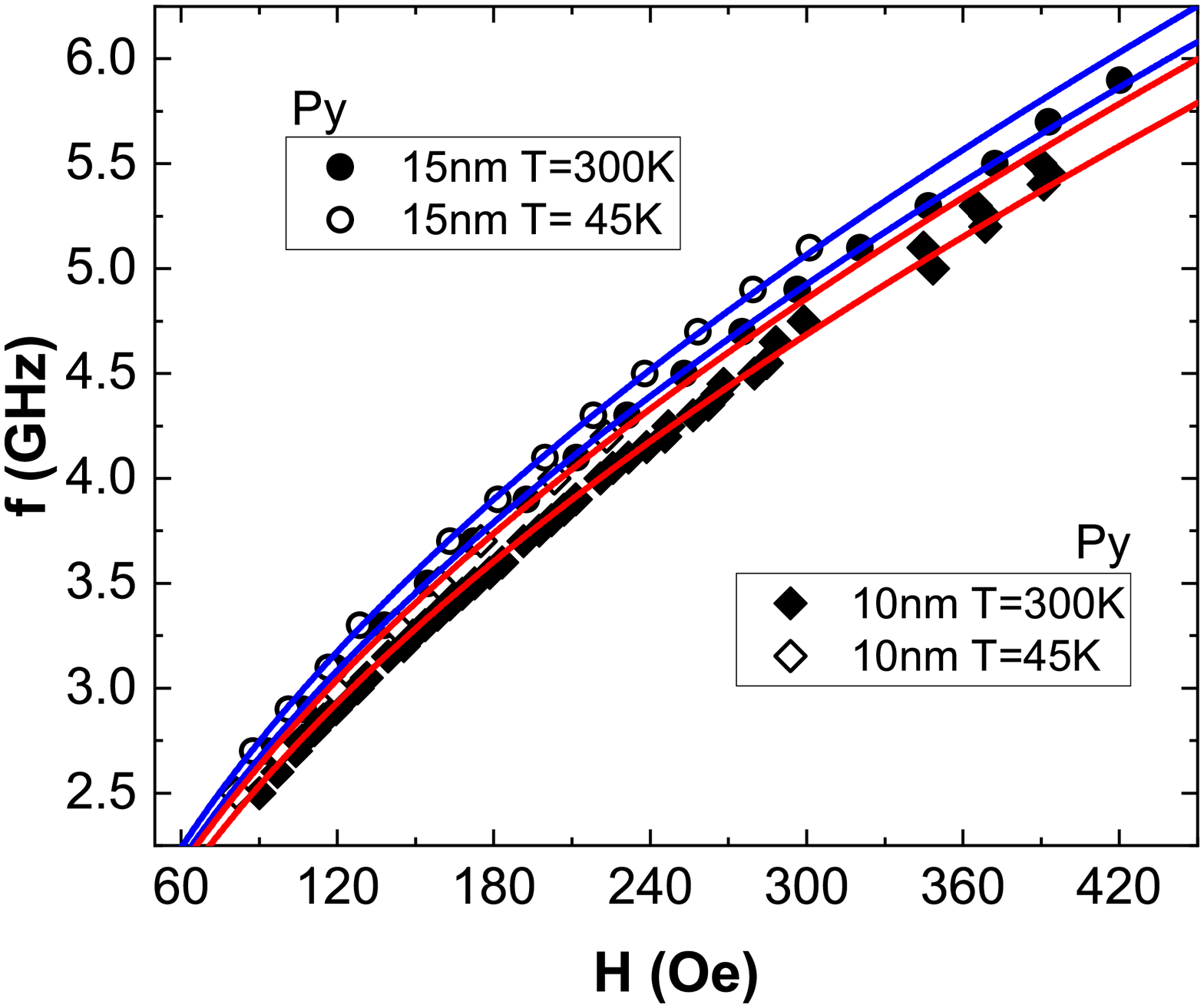}}
\subfigure[]{\includegraphics[width=8cm,height=6cm]{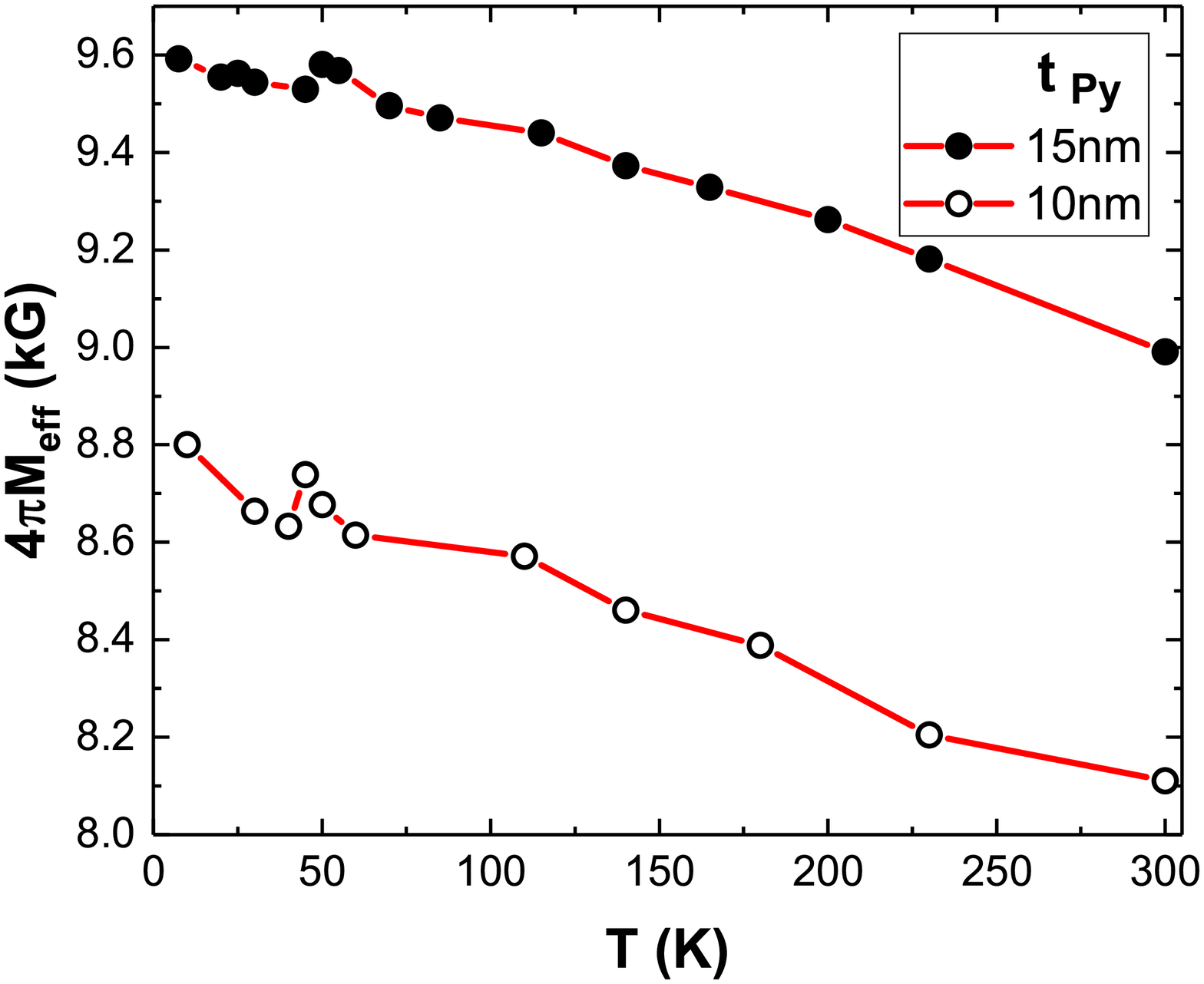}}
\caption{a)Field linewidth variation with resonance frequencies at 300K and 45K for 10 nm and 15nm Py films. Equation 1 has been used for fitting the curve and to determine the Gilbert damping coefficient; b)temperature dependence of damping coefficient for 10nm and 15nm Py thin films; c)resonance field variation with resonance frequencies at 300K and 45K for 10nm and 15nm Py thin films. Kittel formula (eqn-3) has been used for fitting the curve and to determine the $4\pi M_{eff}$; d)temperature dependence of $4\pi M_{eff}$ for 10nm and 15nm Py thin films.}
\label{Fig 5}
\end{figure*}
Fig 4 exhibits the absorption spectra for 15 nm bare Py film after subtraction of a constant background for four different frequencies, 2.5 GHz, 3.5 GHz, 4.5 GHz and 5.5 GHz at room temperature in terms of S-parameter reflection coefficient ($S_{11}$) vs. external magnetic field. We fitted these experimental results to the Lorentz equation \cite{Lorentz Fit}. We extracted the field linewidth  at half maxima from the FMR spectra at different frequencies and fitted them using equation 1 to obtain $\alpha$ as one can see from fig 5a and fig 6a. The experimental values of the absorption linewidth ($\Delta H$) contains both the effect of intrinsic Gilbert damping and the extrinsic contribution to the damping. Linewidth due to Gilbert damping is directly proportional to the resonance frequency and follows the equation:
\begin{equation}
\Delta H=(\frac{2\pi}{\gamma})\alpha f+\Delta H_0
\end{equation} 
where $\gamma$ is the gyromagnetic ratio, $\alpha$ is the Gilbert damping coefficient and $\Delta H_0$ is the inhomogeneous linewidth. A number of extrinsic contributions to the damping coefficient like magnetic inhomogeneities, surface roughness, defects of the thin films bring about the inhomogeneous linewidth broadening \cite{Inhomo}. $\alpha$ has been determined using the above equation only. Damping coefficient values obtained here are in the range of about $0.005$ to $0.009$ for Py samples of thicknesses covering the whole thin film region i.e., 10nm to 90nm at room temperature. These values are pretty close to the values previously reported in the literature \cite{G1,G2,Experimantal_investigation,G3_after review,G4_after review}. For the Co film of thickness 30 nm we have obtained the value of $\alpha$ to be 0.008$\pm$0.0004. Barati \textit{et al.} measured the damping value of 30nm Co film to be 0.004 \cite{CoDamping,CoD2}. There are other literature also where Co multilayers have been studied where damping coefficient value increases because of spin pumping effect.
$\alpha$ is a very interesting parameter to investigate because it is used in the phenomenological LLG equation \cite{Landau}, \cite{Gilbert} to describe magnetization relaxation:
\begin{equation}
\frac{d\vec{M}}{dt}=-\gamma \vec{M} \times \vec{H_{eff}} + \frac{\alpha}{M_S} \vec{M} \times \frac{d \vec{M}}{dt}
\end{equation}
where, $\mu_B$ is Bohr magneton, $\vec{M}$ is the magnetization vector, $M_S$ is the saturation magnetization and $H_{eff}$ is the effectve magnetic field which includes the external field, demagnetization and crystalline anisotropy field. The introduction of the Damping coefficient in LLG equation is phenomenological in nature and the question of whether it has a physical origin or not has not been fully understood till date. We have measured $4\pi M_{eff}$ also from the absorption spectra. We have fitted the Kittel formula (equation 3) into resonance field vs. the resonance frequency ($f_{res}$) data as shown in fig 5c and fig 6c.
\begin{equation}
f_{res}=(\frac{\gamma}{2\pi})[(H+4\pi M_{eff})H]^\frac{1}{2}
\end{equation}
where, $H$ is the applied magnetic field, and $M_{eff}$ is the effective magnetization which contains saturation magnetization and other anisotropic contributions. We obtained the $4\pi M_{eff}$ value for 30nm thick Co and  34nm Py to be 17.4$\pm$0.2kG and 9.6$\pm$0.09kG respectively at room temperature. These values also agree quite well with the literature. For a 10nm Co film, Beaujour \textit{et al.} measured the value to be around $16kG$ \cite{CoM} and for a 30nm Py the value is $10.4kG$ as measured by Zhao \textit{et al}\cite{Experimantal_investigation}.    

We tried to address here the thickness and temperature dependence of $\alpha$ and $4\pi M_{eff}$ using our measurement set-up. 
The variation of the $\alpha$ with thickness is shown here in figure 5b. It increases smoothly as film thickness decreases and then shows a sudden jump below 15nm. Increased surface scattering could be the reason behind this enhanced damping for thinner films. It has been previously observed \cite{Role} that damping coefficient and electrical resistivity follows a linear relation at room temperature for Py thin film. It suggests a strong correlation between magnetization relaxation($\alpha$) and electron scattering. Magnetization relaxation could be explained by electron scattering by phonons and magnons. In the former case, $\alpha$ is proportional to the electron scattering rate, $\tau^{-1}$ and in the later case, $\alpha\sim\tau$. Theoretical predictions by Kambersky \cite{Kambersky} suggests that at higher temperature $\alpha\sim\tau^{-1}$ as electron scattering by phonons are predominant there. So, here in our case we can eliminate the possibility of electron scattering by magnons as thickness dependent study has only been done at room temperature where phonon scattering is prevalent. Ingvasson et.al in their paper\cite{Role} also suggests that the relaxation of magnetization is similar to bulk relaxation where phonon scattering in bulk is replaced by surface and defect scattering in thin films.\\
Thickness dependent study of $4\pi M_{eff}$ also has been done for Py thin films at room temperature. As we can see from fig 5d, $M_{eff}$ is linear for thinner films and becomes almost independent of thickness for thicker films.  The change in $M_{eff}$ with thickness mainly comes from the surface anisotropy,
\begin{equation}
\mu_0 M_{eff}=\mu_0 M_{s}-\frac{2K_s}{M_s d}
\end{equation}
where $M_s$ is the saturation magnetization and $\frac{2K_s}{M_s d}$ is the surface anisotropy field. Surface anisotropy is higher for thinner films and the anisotropy reduces as one increases the film thickness. We have obtained saturation magnetization($4\pi M_{s}$) value of Py to be $10.86 kG$ using the linear fit (equation 4). Previously Chen \textit{et al.} has reported the $4\pi M_{eff}$ value for a 30nm Py film to be $12 kG$ \cite{TM} which includes both $4\pi M_{s}$ and anisotropy field.
\begin{figure}
\centering
\includegraphics[width=8.5cm,height=7cm]{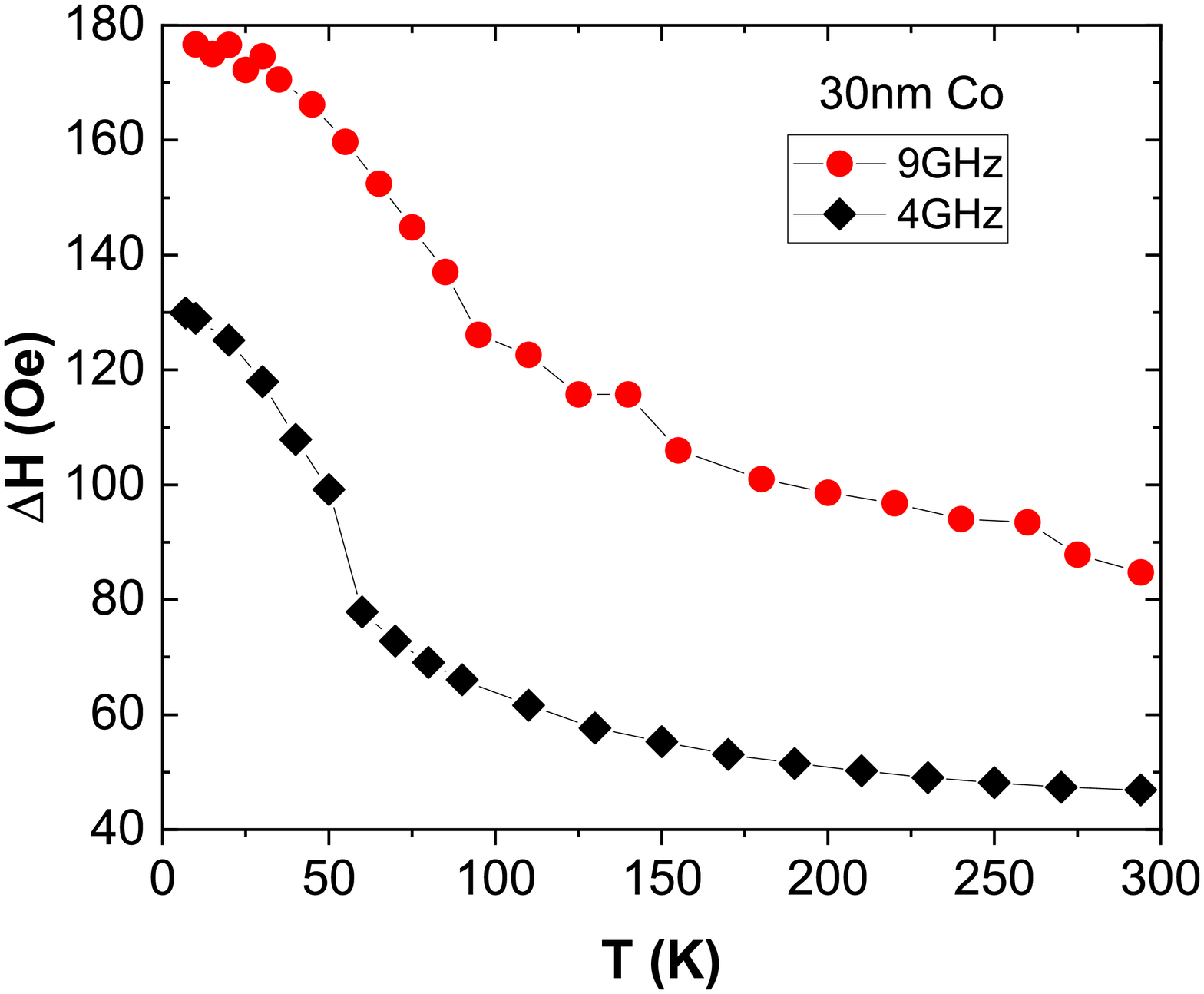}
\caption{Temperature induced linewidth variation of 30nm Co thin film at two different frequencies 4GHz and 9GHz}
\label{Fig 7}
\end{figure}

Temperature dependence of $\alpha$ for 15nm and 10nm Py film is represented in figure 6b. The $\alpha$ value decreases monotonically from room temperature value and reaches a minimum value at around 100K and then starts to increase with further decrease of temperature and reaches a maximum value at 45K. Zhao \textit{et al.} have seen this kind of damping enhancement at around 50K in their low temperature experiment with Py thin films with different types of capping layers and Rio \textit{et al.} observed the damping anomaly at temperature 25K when they have used $Pt$ as a capping layer on Py thin film.\cite{Experimantal_investigation, G1}. We did not use any capping layer on Py film in our measurement. So there is no question of interface effect for the enhanced damping at 45K. A possible reason for the strong enhancement of damping at 45K could be the spin reorientation transition(SRT) on the Py surface at that particular temperature \cite{Experimantal_investigation, Sierra}. Previously it has been established that the competition between different anisotropy energies: magnetocrystalline anisotropy, surface anisotropy, shape anisotropy decides the magnetization direction in magnetic films. For thin films, the variation of temperature, film thickness, strain can alter the competition between shape and surface anisotropy. In our case, temperature variation could be the reason for the spin reorientation transition on Py surface at around 45K.   
For a deeper understanding of the spin reorientation we investigated the temperature dependence of 4$\pi M_{eff}$ for 15nm and 10nm Py film as shown in fig 6d. There $M_{eff}$ is showing an anomaly at around 45K, otherwise it is increasing smoothly with the decrease of temperature. Since there is no reason of sudden change in saturation magnetization at this temperature, the possible reason for the anomaly in $M_{eff}$ should come from any change in magnetic anisotropy. That change of anisotropy can be related to a spin reorientation at that particular temperature value. Sierra $et. al.$, \cite{Sierra}, have also argued that in the temperature dependent spin re-orientation (T-SRT), the central effect of temperature on the magnetic properties of Py films was to increase the in-plane uniaxial anisotropy and to induce a surface anisotropy which orients the magnetization out of plane in the Py surface. They have verified this using X-Ray diffraction experiments and high resolution transmission electron microscopy images. This establishes reasonably enough that it is a spin re-orientation transition around 45K.  

Lastly, for a 30nm Co thin film we have studied the temperature variation of FMR linewidth($\Delta H$) at microwave frequencies 9GHz and 4GHz. One can see from fig7 that the linewidth does not change much in the temperature range 100$<$T$<$300 but below 100K, $\Delta H$ starts to increase significantly. This behaviour of $\Delta H$ has been observed previously by Bhagat \emph{et al.} \cite{Co}. They suggested that the increase of relaxation frequency at low temp might be related to the rapid variation in the electronic mean free path. This comparison suggests that our set-up with a single port CPW has the same sensitivity of the previously reported set-up. 



\section*{CONCLUSIONS}
Short-circuited coplanar waveguide can successfully be used for low-temperature single-port broadband FMR spectroscopy measurements. The magnetic parameters obtained here for the standard ferromagnetic materials, Py and Co are in good agreement with other experimental works and theoretical predictions. Temperature dependent studies of $\alpha$ and $4\pi M_{eff}$ for Py films here exhibit spin reorientation phenomenon at low temperatures. We believe that the findings with Py sample will help in better understanding of magnetic phenomena for other ferromagnetic materials at low temperature. Though this setup has limitations for angle-dependent FMR measurements, it can readily be used for studies on magnetization dynamics for multi-layered films and planar nanostructures and is currently under
way. In future we aim to integrate electrical measurement facilities with the existing set-up thereby extending the possibilities of inverse spin hall(ISHE) measurements and spin-torque ferromagnetic resonance(ST-FMR) spectroscopy experiments which are very relevant for current scientific interests. We believe the short-circuited CPW will serve its purpose conveniently there too.

\section*{ACKNOWLEDGEMENTS}
The authors sincerely acknowledges Ministry of Education, Government of India and Science and Engineering Research Board (SERB) (grant no:EMR/2016/007950) and Department of Science and Technology (grant no. DST/ICPS/Quest/2019/22) for financial support. S.P. acknowledges Department of Science and Technology(DST)-INSPIRE fellowship India, S. A. acknowledges  Ministry of Education of Government of India and S.M. acknowledges Council Of Scientific and Industrial Research(CSIR),India for research fellowship. The authors would like to thank Dr. Partha Mitra of the Department of Physical Sciences, Indian Institute of Science Education and Research Kolkata for providing the lab facilities for sample deposition. The authors would also like to thank Mr. Subhadip Roy, IISER Kolkata for his help in fabricating the CPW structure using home-built optical lithography set-up.

\end{document}